\documentclass[twocolumn,twoside,slac_two]{revtex4}
\usepackage{graphicx}
\usepackage{fancyhdr}
\begin{document}

\title{$\phi_1/\beta$ from $B \to charmonium/charm$ Modes}

%

\author{T. Hara}
\affiliation{Osaka University, Toyonaka, Osaka 560-0043, Japan}

\begin{abstract}
The asymmetric B-factories have provided 
valuable information on $CP$ violation so far. 
In particular, one of the angles of the Unitarity
Triangle, $\phi_1(=\beta)$, has been measured by several approaches.
Since FPCP2004, some measurements have been updated and improved.
In this Letter, the latest status of $\phi_1$ measurements,
performed at BaBar and Belle experiments using $B \to charm/charmonium$ decays,
are reported.
\end{abstract}

\maketitle

\thispagestyle{fancy}


\section{Introduction}

$CP$ violation in the Standard Model (SM) stems from 
an irreducible complex phase in the weak interaction 
$3 \times 3$ quark-mixing (CKM) matrix~\cite{KM}. 
Applying the unitarity constraint, especially between the first and third generation, 
an equation $V^{}_{ud}V^*_{ub}+V^{}_{cd}V^*_{cb}+V^{}_{td}V^*_{tb}=0$, 
which can be depicted as an Unitarity 
Triangle in a complex plane, is required to be satisfied. 
The key objective of B-factories~\cite{BaBar-Det, Belle-Det} 
is to determine three angles or 
sides of this triangle.
 
At the B-factories, it is predicted that a $CP$ violating asymmetry 
lies in the time-dependent rates for $B^0$ and $\bar{B}^0$
decays to a common $CP$ eigenstate ``$f_{CP}$'',
which is generally written as:
\begin{eqnarray}
A(t) & \equiv & \frac{\Gamma(\bar{B}^0\to f_{CP})-\Gamma({B^0}\to f_{CP})}
{\Gamma(\bar{B}^0\to f_{CP})+\Gamma({B^0}\to f_{CP})}  \nonumber \\
& = & S_{f_{cp}}\sin\Delta m_d t + A_{f_{cp}}\cos\Delta m_d t \nonumber
\label{eq-sp}
\end{eqnarray}

\[
 S_{f_{cp}}=\frac{2 Im\lambda}{|\lambda|^2+1}\hspace{2mm}{\rm and}\hspace{2mm}
 A_{f_{cp}}=\frac{|\lambda|^2-1}{|\lambda|^2+1}
\]
\noindent
where $\Gamma(\bar{B}^0$ $(B^0)$
$\to f_{CP})$ is the decay rate
for a $\bar{B}^0(B^0)$ to $f_{CP}$ at a proper time $t$ after production,
$\Delta m_d$ is the mass difference between the two $B^0$ mass eigenstates,
$A_{f_{cp}}$ \footnote{sometimes, $A_{f_{cp}}$ is expressed as 
$-C_{f_{cp}}$.} and $S_{f_{cp}}$ are expressed with a complex parameter, $\lambda$,
that depends both on $\bar{B}^0-{B^0}$ mixing and on the 
amplitudes for $\bar{B}^0 ({B^0})$ decay to $CP$ eigenstates.
In the SM, $|\lambda|$ is equal to the absolute value of the ratio of the
$\bar{B}^0$ to ${B^0}$ decay amplitude to a good approximation.

In this Letter, the latest status of $\phi_1$ measurements are reported,
here $\phi_1$ is the one of the angles of the Unitarity Triangle,
performed at BaBar and Belle experiments using $B \to charm/charmonium$ 
\footnote {Unless explicitly stated, charge conjugate decay modes are
assumed throughout this letter.} decays.

%

\section{$b \to c \bar{c} s$ Decay Modes}
\begin{figure}[h]
\centering
\includegraphics[width=80mm]{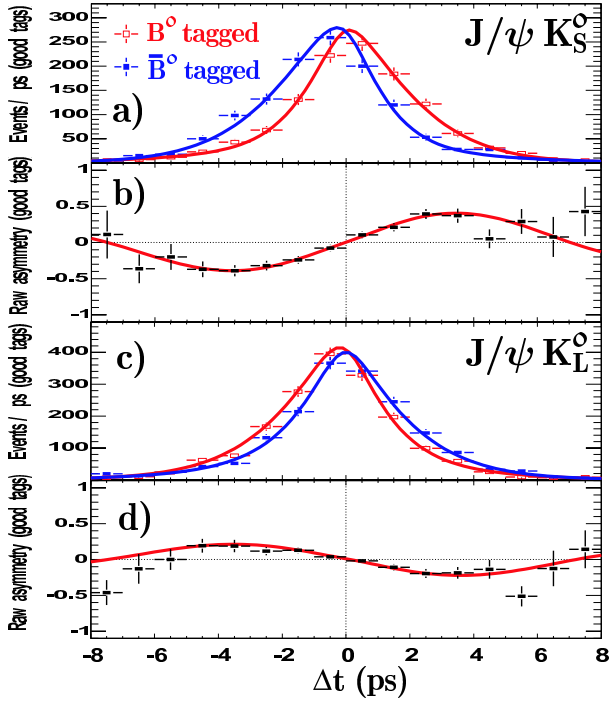}
\caption{a(c) shows the observed proper time distribution for
$B^0 \to J/\psi K_S$($J/\psi K_L$) at Belle. Open(closed) squares represent events
tagged as $B^0(\bar{B}^0)$. 
b(d) shows the corresponding raw asymmetry together with an
unbinned maximum-likelihood fit result.} \label{belle-sin2phi1}
\end{figure}

The determination of $\phi_1$ from $b \to c \bar{c} s$ decay modes,
as typified by $B^0 \to J/\psi K^0$,
provides the stringent constraint on the Unitarity Triangle at this stage.
For these decay modes, the $CP$ violation parameters are rather 
precisely expressed as
$S_{b \to c \bar{c} s}=-\xi_f\sin2\phi_1$ and $A_{b \to c \bar{c} s}=0$, where
$\xi_f$ is $-1$ for $(c\bar c) K_S$ and $+1$ for $(c\bar c) K_L$ final state.
Using $B^0 \to J/\psi K^0$ decays recorded in 386M $B \bar{B}$ events,
Belle has updated the results~\cite{Belle-jpsiks}.
Figure~\ref{belle-sin2phi1}a(c) and b(d) show the observed proper time distribution for
$B^0 \to J/\psi K_S$($J/\psi K_L$) and corresponding raw asymmetry.
The $CP$ violation parameters are determined by performing an
unbinned maximum-likelihood fit to the proper time distributions.
The $K_S$ and $K_L$ samples are combined with taking into account the $CP$ eigenstates.
The fit results are $S_{b \to c \bar{c} s}(= \sin2\phi_1) = 
0.652 \pm 0.039({\rm stat}) \pm 0.020({\rm syst})$
and $A_{b \to c \bar{c} s} = 0.010 \pm 0.026 \pm 0.036$.
BaBar has also measured these parameters with considering not only $J/\psi K^0$
but also $\psi(2S) K_S$, $\chi_{c1} K_S$ and $\eta_c K_S$ decays observed in 227M
$B \bar{B}$ events~\cite{BaBar-jpsiks} and determined to be,
$\sin2\phi_1 = 0.722 \pm 0.040 \pm 0.023$ and 
$A_{b \to c \bar{c} s} = -0.051 \pm 0.033 \pm 0.014$.

\begin{figure}[h]
\centering
\includegraphics[width=80mm]{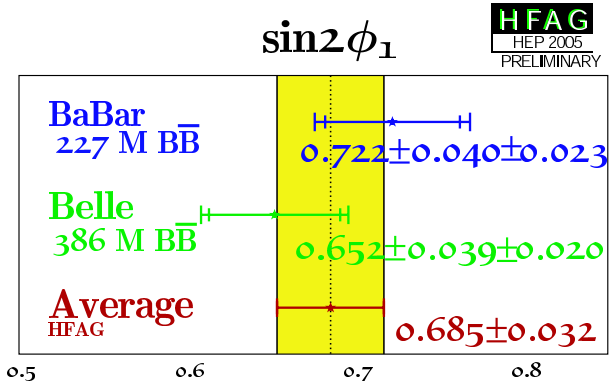}
\caption{$\sin 2\phi_1$ from B-factories.} \label{wa-sin2phi1}
\end{figure}

The combined results from B-factories~\cite{HFAG-jpsiks} (summarized in Fig.~\ref{wa-sin2phi1}) are
\begin{eqnarray}
 S_{b \to c \bar{c} s} & = \sin2\phi_1 & = 0.685 \pm 0.032(0.028 {\rm stat. only}) \nonumber \\
 A_{b \to c \bar{c} s} &               & =-0.026 \pm 0.041(0.020 {\rm stat. only}) \nonumber
\end{eqnarray}

\noindent
and show a good agreement with the SM expectations.
The relative error for $\sin 2\phi_1$ has reached less than 5\% level.
This fact allows us to test a new physics effect which would manifests itself
in a loop-diagram such as $b \to s$ penguin~\cite{BaBar-phiks, BaBar-pi0ks, Belle-jpsiks}.

\section{$\cos 2\phi_1$ Measurements}

The analyses for $b \to c \bar{c} s$ decay modes impose constraints 
on $\sin 2\phi_1$ only and then lead to a four-fold ambiguity in the
determination of $\phi_1$.
The next natural step is to reduce this ambiguity from four-fold
to two-fold by measuring the sign of $\cos 2\phi_1$.
For this purpose, the following two methods have been performed.

\subsection{$B^0 \to J/\psi K^{*0}$ Decay}

Because the $B^0 \to J/\psi K^{*0}$ mode is a pseudoscalar decay to
two vector particles, interference between three helicity states with 
different $CP$ eigenstates contributes to the 
$CP$ violation parameters $\sin 2\phi_1$ and $\cos 2\phi_1$~\cite{Theo1-jpsikst0,
Theo2-jpsikst0, Theo3-jpsikst0, Theo4-jpsikst0}.
To extract these, the complex transversity amplitudes of three helicity states and 
strong phase differences are necessary to be measured in advance.
BaBar~\cite{BaBar-jpsikst0} and Belle~\cite{Belle-jpsikst0} have performed this analysis.
In particular, BaBar has measured these through a time-integrated angular 
analysis of flavor specific decays, such as 
$B^+ \to J/\psi K^{*+} (K^{*+} \to K_S \pi^+ {\hspace{1mm}\rm{and}\hspace{1mm}}
K^+ \pi^0)$ and $J/\psi K^{*0} (K^{*0} \to K^+ \pi^-)$ modes recorded 
in 88M $B \bar{B}$ events,
considering the interference between S-wave and P-wave $K\pi$ 
final states~\cite{BaBar-jpsikst0}.
Using these values, then an unbinned maximum likelihood fit has been applied to 
the time-dependent angular distribution of the $CP$ decay mode
$B^0 \to J/\psi K^{*0} (K^{*0} \to K_S^0 \pi^0)$.
The extracted $\sin2\phi_1$ and $\cos2\phi_1$ from this fit are;
\begin{eqnarray}
 \sin2\phi_1 & = & -0.10 \pm 0.57 \pm 0.14 \nonumber \\
 \cos2\phi_1 & = &  3.32^{+0.76}_{-0.96} \pm 0.27 \nonumber
\end{eqnarray}
\noindent
When $\sin2\phi_1$ is fixed to 0.731
(this was the world average as of this analysis), the resultant
$\cos2\phi_1 = 2.72^{+0.50}_{-0.79} \pm 0.27$ is obtained and 
the sign of $\cos2\phi_1$ is found to be positive at the 86\% C.L.

\subsection{$B^0 \to D[K_S^0 \pi^+ \pi^-] h^0$ Decay}

Another technique of reducing $\phi_1$ ambiguity 
is based on the analysis of $B^0 \to D_{CP} h^0$
governed by the CKM favored $b \to c \bar{u} d$ transitions~\cite{Theo-dalitz}.
Here $h^0$ means light neutral mesons.
Though the CKM suppressed $b \to u \bar{c} d$ decays contribute
to these modes, the effect is small and can be taken into account.
When $D_{CP}$ decays to $K_S^0 \pi^+ \pi^-$, a time dependent Dalitz 
plot analysis of the neutral $D$ decay allows to determine $\phi_1$
directly.

Belle has performed an analysis for the decays $B \to D \pi^0$,
$D \eta$, $D \omega$ and also for the $B \to D^* \pi^0$, $D^* \eta$
($D^* \to D \pi^0$ and $D \to K_S^0 \pi^0 \pi^-$)
collected in 386M $B \bar{B}$ events~\cite{Belle-dalitz}.
Totally $309\pm31$ events were observed.
An unbinned maximum likelihood fit has been applied to the
observed time-dependent Dalitz plot.
The results are given in Table~\ref{dalitz-table} for each of the
three final states separately.
Then the result of the simultaneous fit over all these modes is;

\[
\phi_1 = 16\pm21\pm11 {}^\circ \nonumber
\]

The first error is statistical and the second is systematic.
The 95\% C.L. region including systematic uncertainty is 
$-30^\circ < \phi_1 < 62^\circ$.

\begin{table}[h]
\begin{center}
\caption{Fit results of $\phi_1$ for the three final states.
The errors are statistical only.}
\begin{tabular}{|l|c|}
\hline \textbf{Final state} & \textbf{$\phi_1 [{}^\circ]$} 
\\
\hline $D \pi^0$  & $11\pm26$ \\
\hline $D \omega$, $D \eta$ & $28\pm32$ \\
\hline $D^* \pi^0, D^* \eta$ & $25\pm35$ \\
\hline
\end{tabular}
\label{dalitz-table}
\end{center}
\end{table}

\begin{figure}[h]
\centering
\includegraphics[width=80mm]{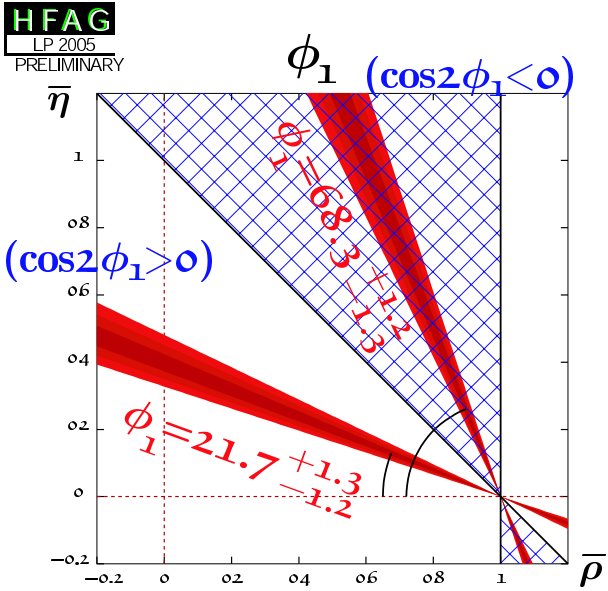}
\caption{Constraint for $\phi_1$ on the ($\bar{\rho},
\bar{\eta}$) plane, obtained from the analysis of $b \to c \bar{c} s$ 
decays, the angular analysis of $B \to J/\psi K^{*0}$ and 
the time-dependent Dalitz plot analysis of $B \to D_{CP} h^0$.
The hatched area, corresponding to the solutions with negative
$\cos2\phi_1$, is ruled out at $\sim 3\sigma$.} 
\label{phi1-all}
\end{figure}

Taking into account both the angular analysis of $B \to J/\psi K^{*0}$ and
the time-dependent Dalitz plot analysis of $B \to D_{CP} h^0$,
the solutions with negative $\cos 2\phi_1$ can be considered to be 
ruled out at $\sim 3\sigma$~\cite{HFAG-jpsiks}.
The resultant constraint to the $\sin 2\phi_1$ obtained by these 
$\phi_1$ measurements is shown in Fig.~\ref{phi1-all}.

\section{$b \to c \bar{c} d$ Decay Modes}
The comparison $\sin2\phi_1$ derived from $b \to c \bar{c} s$ modes
with that from different quark transition is an important
test to check the SM. 
Because a sizable $CP$ asymmetry deviated from the SM expectation 
might be observed, if new physics beyond the SM exists.
The decays dominated by the $b \to c \bar{c} d$ transition,
such as $B^0 \to J/\psi \pi^0$ and $D^{(*)+}D^{(*)-}$, 
are suitable for this check~\cite{BaBar-jpsipi0, Belle-jpsipi0,
BaBar-D+D-, BaBar-DstDst, Belle-DstDst, Belle-DstD}.

Since the FPCP2004, BaBar has updated the results for these decay modes.
Recently, the improved measurement for $B^0 \to J/\psi \pi^0$ mode 
observed in 232M $B \bar{B}$ events has been reported~\cite{BaBar-jpsipi0}.
The $109\pm12$ signal events are observed and $CP$ violation parameters
are determined to be,
\begin{eqnarray}
 S_{J/\psi \pi^0} & = & 0.68 \pm 0.30 \pm 0.04 \nonumber \\
 C_{J/\psi \pi^0} (= -A_{J/\psi \pi^0}) & = &-0.21 \pm 0.26 \pm 0.06 \nonumber
\end{eqnarray}
These are consistent both with previous measurements from B-factories~\cite{Belle-jpsipi0}
and with the SM expectations.

\begin{figure}[h]
\centering
\includegraphics[width=80mm]{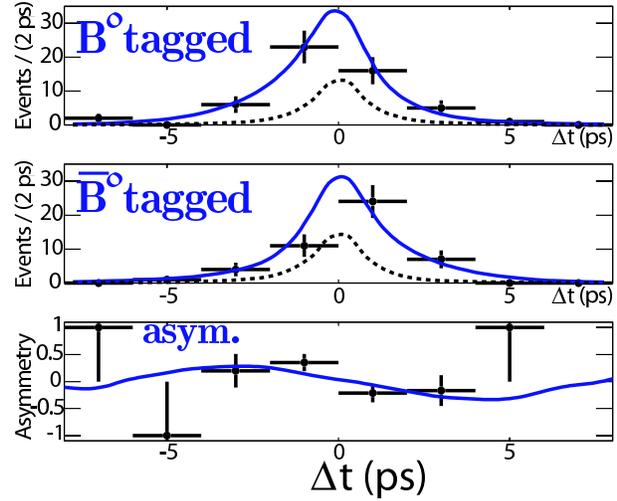}
\caption{The proper time distribution for $B \to J/\psi \pi^0$ tagged
as $B^0$ (top) and $\bar{B}^0$ (middle) observed at BaBar. 
The solid lines are the
sum of signal and backgrounds and the dotted lines show the background 
components. The time-dependent $CP$ asymmetry is shown (bottom) with an
unbinned maximum-likelihood fit result.} \label{babar-jpsipi0}
\end{figure}

Other improved measurements for the $B^0 \to D^{(*)+}D^{-}$ decays
have been also reported.
In particular, BaBar has performed a first measurement of $CP$ asymmetries
in the $B^0 \to D^+ D^-$ decay~\cite{BaBar-D+D-}.
The measured values are 
\begin{eqnarray}
 S_{D^+ D^-} & = &-0.29 \pm 0.63 \pm 0.06 \nonumber \\
 C_{D^+ D^-} & = &-0.11 \pm 0.35 \pm 0.06 \nonumber
\end{eqnarray}
and consistent again with the SM expectations within the 
currently large statistical uncertainty, that will be
more precise in the future.

\section{Summary}
One of the angles of the Unitarity
Triangle, $\phi_1(=\beta)$, has been measured by several approaches
at B-factories, BaBar and Belle experiments.
The analyses for $b \to c \bar{c} s$ decay modes impose
the stringent constraint on $\sin 2\phi_1$ with a precision of
$\sim 5\%$.

Then the four-fold ambiguity of $\phi_1$, which are allowed 
in $[0, 2\pi]$ mathematically, are reduced
by the time-dependent angular analysis of $B^0 \to J/\psi K^{*0}$
and the time-dependent Dalitz plot analysis of $D_{CP}$ in 
$B^0 \to D_{CP} h^0$ decays, 
the solutions with negative $\cos 2\phi_1$ can be considered to be
ruled out at $\sim 3\sigma$.

The comparisons $\sin2\phi_1$ derived from $b \to c \bar{c} s$ modes
with that from different quark transition, $b \to c \bar{c} s$, also have been
done, because a sizable $CP$ asymmetry deviated from the SM expectation 
might be observed, if new physics beyond the SM exists.
Though these measurements are dominated by statistical uncertainties,
more significant measurements will be anticipated in the future.

\begin{acknowledgments}
It is pleasure to thank G.~Cavoto and K.~Trabelsi for providing information
on the status of $\phi_1/\beta$ measurements performed at BaBar and Belle.
I wish to thank the all organizers for giving me an opportunity to 
take part in the conference.
\end{acknowledgments}

\bigskip 

\begin{thebibliography}{99} 

\bibitem{KM}M. Kobayashi and T. Maskawa, {\em Prog. Theor. Phys.} {\bf 49}, 652 (1973).
\bibitem{BaBar-Det}B. Aubert {\it et al.}, {\em Nucl. Instrum. Meth.} {\bf A 479}, 1 (2003).
\bibitem{Belle-Det}A. Abashian {\it et al.}, {\em Nucl. Instrum. Meth.} {\bf A 479}, 117 (2003).
\bibitem{Belle-jpsiks}K. Abe {\it et al.}, arXiv:hep-ex/0507037.
\bibitem{BaBar-jpsiks}B. Aubert {\it et al.}, {\em Phys. Rev. Lett.} {\bf 94}, 161803 (2005).
\bibitem{HFAG-jpsiks} http://www.slac.stanford.edu/xorg/hfag/triangle /moriond2006/index.shtml
\bibitem{BaBar-phiks}B. Aubert {\it et al.}, {\em Phys. Rev.} {\bf D 71}, 091102(R) (2005).
\bibitem{BaBar-pi0ks}B. Aubert {\it et al.}, {\em Phys. Rev.} {\bf D 71}, 111102 (2005).
\bibitem{Theo1-jpsikst0}I. Dunietz, H. Quinn, A. Snyder, W. Toki and H. J. Lipkin, {\em Phys. Rev.} {\bf D 43}, 2193 (1991).
\bibitem{Theo2-jpsikst0}J. Charles, A. Le. Yaouanc, L. Oliver, O. Pene and J. C. Raynal, {\em Phys. Rev.} {\bf D 58}, 114021 (1998).
\bibitem{Theo3-jpsikst0}A. S. Dighe, I. Dunietz and R. Fleischer, {\em Phys. Lett.} {\bf B 433}, 147 (1998).
\bibitem{Theo4-jpsikst0}C. W. Chiang, {\em Phys. Rev.} {\bf D 62}, 014017 (2000).
\bibitem{BaBar-jpsikst0}B. Aubert {\it et al.}, {\em Phys. Rev.} {\bf D 71}, 032005 (2005).
\bibitem{Belle-jpsikst0}R. Itoh, Y. Onuki {\it et al.}, {\em Phys. Rev. Lett.} {\bf 95}, 091601 (2005).
\bibitem{Theo-dalitz}A. Bonder, T. Gershon and P. Krokovny, {\em Phys. Lett.} {\bf B 624}, 1 (2005).
\bibitem{Belle-dalitz}K. Abe {\it et al.}, arXiv:hep-ex/0507065.

\bibitem{BaBar-jpsipi0}B. Aubert {\it et al.}, arXiv:hep-ex/0603012.
\bibitem{Belle-jpsipi0}S. U. Kataoka {\it et al.}, {\em Phys. Rev. Lett.} {\bf 93}, 261801 (2004).
\bibitem{BaBar-D+D-}B. Aubert {\it et al.}, {\em Phys. Rev. Lett.} {\bf 95}, 131802 (2005).
\bibitem{BaBar-DstDst}B. Aubert {\it et al.}, {\em Phys. Rev. Lett.} {\bf 95}, 151804 (2005).
\bibitem{Belle-DstDst}H. Miyake, M. Hazumi {\it et al.}, {\em Phys. Lett.} {\bf B 618}, 34 (2005).
\bibitem{Belle-DstD}T. Aushev {\it et al.}, {\em Phys. Rev. Lett.} {\bf 93}, 201802 (2004).

\end{thebibliography}

\end{document}